\documentclass[conference]{IEEEtran}
\usepackage[letterpaper, left=0.56in, right=0.56in, bottom=1in, top=0.7in]{geometry}
\IEEEoverridecommandlockouts

\usepackage[acronym]{glossaries}
\newacronym{BS}{BS}{base station}
\newacronym{PS}{PS}{phase-shifter}
\newacronym{RL}{RL}{reinforcement learning}
\newacronym{AP}{AP}{analog precoder}
\newacronym{FC-HBF}{FC-HBF}{fully-connected HBF}
\newacronym{FSA-HBF}{FSA-HBF}{fixed subarray HBF}
\newacronym{DSA-HBF}{DSA-HBF}{dynamic subarray HBF}
\newacronym{BF}{BF}{beamforming}
\newacronym{UE}{UE}{user equipment}
\newacronym{AWGN}{AWGN}{additive white gaussian noise}
\newacronym{MIMO}{MIMO}{multiple-input multiple-output}
\newacronym{MISO}{MISO}{multiple-input single-output}
\newacronym{RF}{RF}{radio frequency}
\newacronym{RIS}{RIS}{reconfigurable intelligent surfaces}
\newacronym{IOT}{IOT}{internet-of-things}
\newacronym{CL}{CL}{convolutional layer}
\newacronym{FDD}{FDD}{frequency division duplex}
\newacronym{TDD}{TDD}{time division duplex}
\newacronym{CSI}{CSI}{channel state information}
\newacronym{DNN}{DNN}{deep neural network}
\newacronym{DP}{DP}{digital precoder}
\newacronym{DL}{DL}{deep learning}
\newacronym{SVD}{SVD}{singular-value decomposition}
\newacronym{CNN}{CNN}{convolution neural network}
\newacronym{FDP}{FDP}{fully digital precoder}
\newacronym{SE}{SE}{spectral efficiency}
\newacronym{OFDM}{OFDM}{orthogonal frequency division multiplexing}
\newacronym{OMP}{OMP}{orthogonal matching pursuit}
\newacronym{FL}{FL}{fully-connected layer}
\newacronym{HSHO}{HSHO}{Hybrid Structured Heuristic Optimization}
\newacronym{HBF}{HBF}{hybrid beamforming}
\newacronym{IA}{IA}{initial access}
\newacronym{mm-Wave}{mm-Wave}{millimeter wave}
\newacronym{mMIMO}{mMIMO}{massive MIMO}
\newacronym{SINR}{SINR}{signal-to-interference-noise ratio}
\newacronym{SNR}{SNR}{signal-to-noise ratio}
\newacronym{RSSI}{RSSI}{received signal strength indicator}
\newacronym{PZF}{PZF}{phase zero forcing}
\newacronym{PSO}{PSO}{particle swarm optimization}
\newacronym{ZF}{ZF}{zero forcing}
\newacronym{O-FDP}{O-FDP}{optimal fully digital precoder}
\newacronym{JT}{JT}{joint transmission}
\newacronym{CU}{CU}{central unit}
\newacronym{MSE}{MSE}{mean square error}
\newacronym{CEL}{CEL}{cross entropy loss}
\newacronym{CB}{CB}{conjugate beamforming}
\newacronym{NC}{NC}{network controller}
\newacronym{CoMP}{CoMP}{coordinated multi point}
\newacronym{CF-mMIMO}{CF-mMIMO}{cell-free massive MIMO}
\newacronym{CF-HBF}{CF-HBF}{cell-free hybrid beamforming}
\newacronym{CF-BF}{CF-BF}{cell-free beamforming}
\newacronym{MLDG}{MLDG}{Meta-Learning Domain Generalization}
\newacronym{MAML}{MAML}{Model Agnostic Meta-Learning}
\newacronym{WSR}{WSR}{Weighted Sum Rate}

\newif\ifDeepMIMOModel
\DeepMIMOModeltrue

\newif\ifSimpleNParamEq
\SimpleNParamEqtrue

\usepackage{lettrine}
\usepackage[ruled,vlined,linesnumbered]{algorithm2e}
\usepackage{multirow}
\usepackage{longtable}
\usepackage[table,xcdraw]{xcolor}

\makeatletter
\let\oldlt\longtable
\let\endoldlt\endlongtable
\def\longtable{\@ifnextchar[\longtable@i \longtable@ii}
\def\longtable@i[#1]{\begin{figure}[t]
\onecolumn
\begin{minipage}{0.5\textwidth}
\oldlt[#1]
}
\def\longtable@ii{\begin{figure}[t]
\onecolumn
\begin{minipage}{0.5\textwidth}
\oldlt
}
\def\endlongtable{\endoldlt
\end{minipage}
\twocolumn
\end{figure}}
\makeatother

\usepackage[noend]{algpseudocode}
\usepackage{ifpdf}

\ifCLASSINFOpdf
  \usepackage[pdftex]{graphicx}
  \usepackage{graphicx,epstopdf}
  \graphicspath{{../pdf/}{../jpeg/}}
  \DeclareGraphicsExtensions{.pdf,.jpeg,.png,.eps}
\else
  \usepackage[dvips]{graphicx}
  \usepackage{graphicx,epstopdf}
  \graphicspath{{../eps/}}
  \DeclareGraphicsExtensions{}
\fi
\usepackage{tikz}
\usetikzlibrary{decorations.pathreplacing,calc}
\newcommand{\tikzmark}[1]{\tikz[overlay,remember picture] \node (#1) {};}

\newcommand*{\AddNote}[4]{%
    \begin{tikzpicture}[overlay, remember picture]
        \draw [decoration={brace,amplitude=0.5em},decorate,line width=.2mm,black]
            ($(#3)!(#1.north)!($(#3)-(0,1)$)$) --  
            ($(#3)!(#2.south)!($(#3)-(0,1)$)$)
                node [align=center, text width=2.5cm, pos=0.5, anchor=west] {#4};
    \end{tikzpicture}
}%

\usepackage{cite}
\usepackage{amsthm}
\usepackage{steinmetz}
\usepackage{amssymb}
\usepackage{balance}
\usepackage{eqparbox}
\usepackage{multirow}
\usepackage{bbm}
\usepackage{float}
\SetAlFnt{\small}

\usepackage{amsmath}

\usepackage{mathtools, nccmath}

\usepackage{booktabs, siunitx}
\usepackage[scaled]{DejaVuSansMono}
\usepackage[T1]{fontenc}

\usepackage{color}
\usepackage{relsize}
\usepackage{mathtools}

\newcommand{\bs}[1]{\boldsymbol{#1}}

\newcommand{\mb}[1]{\mathbf{#1}}

\usepackage{mathtools}

\SetKwInput{KwInput}{Input}                
\SetKwInput{KwOutput}{Output}
\SetKwInput{KwOutputr}{Output Regression}              
\SetKwInput{KwOutputc}{Output Classification}              

\newcommand{\bseq}{\begin{subequations}}
\newcommand{\eseq}{\end{subequations}}
\newcommand{\baln}{\begin{align}}
\newcommand{\ealn}{\end{align}}
\newcommand{\balnd}{\begin{aligned}}
\newcommand{\ealnd}{\end{aligned}}
\newcommand{\beq}{\begin{equation}}
\newcommand{\eeq}{\end{equation}}
\newcommand{\beqn}{\begin{eqnarray}}
\newcommand{\eeqn}{\end{eqnarray}}
\newcommand{\beqno}{\begin{eqnarray*}}
\newcommand{\eeqno}{\end{eqnarray*}}
\newcommand{\bma}{\begin{displaymath}}
\newcommand{\ema}{\end{displaymath}}
\newcommand{\bnu}{\begin{enumerate}}
\newcommand{\enu}{\end{enumerate}}
\newcommand{\bce}{\begin{center}}
\newcommand{\ece}{\end{center}}
\newcommand{\btb}{\begin{tabular}}
\newcommand{\etb}{\end{tabular}}
\newcommand{\ba}{\begin{array}}
\newcommand{\ea}{\end{array}}
\usepackage{footnote}
\makesavenoteenv{tabular}
\makesavenoteenv{table}
\makeatletter 
\newcommand\semiHuge{\@setfontsize\semiHuge{21.1}{27.38}}
\makeatother

\begin{document}

\title{SAGE-HB: Swift Adaptation and Generalization in Massive MIMO Hybrid Beamforming
 \thanks{ 
 The authors are with the Department of Electrical Engineering, Polytechnique Montreal, Montreal, QC H3T 1J4, Canada. Emails: \{ali.hasanzadeh-karkan, hamed.hojatian, j-f.frigon, francois.leduc-primeau\}@polymtl.ca.}
}

\author{\IEEEauthorblockN{Ali~Hasanzadeh~Karkan, Hamed~Hojatian, Jean-François~Frigon, and François~Leduc-Primeau\\
}}

\maketitle

\IEEEpubidadjcol

\begin{abstract}
Deep learning~(DL)-based solutions have emerged as promising candidates for beamforming in massive Multiple-Input Multiple-Output~(mMIMO) systems. Nevertheless, it remains challenging to seamlessly adapt these solutions to practical deployment scenarios, typically necessitating extensive data for fine-tuning while grappling with domain adaptation and generalization issues. In response, we propose a novel approach combining Meta-Learning Domain Generalization~(MLDG) with novel data augmentation techniques during fine-tuning. This approach not only accelerates adaptation to new channel environments but also significantly reduces the data requirements for fine-tuning, thereby enhancing the practicality and efficiency of DL-based mMIMO systems.
The proposed approach is validated by simulating the performance of a backbone model when deployed in a new channel environment, and with different antenna configurations, path loss, and base station height parameters. Our proposed approach demonstrates superior zero-shot performance compared to existing methods and also achieves near-optimal performance with significantly fewer fine-tuning data samples.
\end{abstract}

\vspace{-5pt}
\section{Introduction} \label{sec:intro}
The emergence of \gls{mMIMO} technology ushers in a new era of wireless communication, promising unprecedented gains in spectral efficiency and network capacity~\cite{andrews2014what}. \Gls{HBF} offers a balanced solution, merging analog and digital beamforming to efficiently deploy \gls{mMIMO} systems. Unlike \gls{FDP}, \gls{HBF} strikes a practical solution, optimizing signal transmission while overcoming hardware constraints in large-scale deployments~\cite{molisch2017hybrid}.

Alongside traditional beamforming methods, recent advances in \gls{DL}-based \gls{HBF} design, encompassing supervised, unsupervised, and reinforcement learning paradigms, have drawn significant attention~\cite{huang2018deep, alkhateeb2018deep, hojatian2021unsupervised, elbir2020hybrid, hojatian2022flexible, wang2020precodernet}. These DL-based HBF methods not only hold the potential for improved spectral efficiency but also offer the promise of lower complexity. Supervised learning excels in precision but lacks adaptability and relies on labeled data, while unsupervised learning is more adaptive than supervised learning but is not known for its fast adaptability; it often involves time-complex fine-tuning that can be ineffective. Reinforcement learning holds promise for adaptability in dynamic environments but imposes significant computational complexity and maintaining adaptability, requiring extensive resources and potentially impractical training times.
In practical deployment scenarios, the need for fine-tuning poses a significant challenge for \gls{DL}-based approaches. 
The data collection not only incurs resource costs but, more critically, reduces the base station's operational efficiency. There is thus a strong need for resource-efficient fine-tuning in DL implementation.

Recent studies have explored transfer learning, and meta-learning techniques aimed at enhancing the adaptation of pre-trained DNN to the specific requirements of deployment \gls{BS} \cite{yuan2020transfer, zhang2021embedding, xia2021meta}. 
Particularly, in \cite{yuan2020transfer}, the authors propose two offline adaptive beamforming algorithms based on deep transfer learning and meta-learning, and an online algorithm to enhance the adaptation capability of the offline meta-algorithm in realistic non-stationary environments.
The authors in \cite{zhang2021embedding} presented a solution for tackling the problem of mismatch by first training an embedding model to extract transferable features, and then using support vector regression for fitting.
An iterative meta-learning algorithm is introduced in \cite{xia2021meta} for maximizing \gls{WSR} in MISO downlink channels.
In \cite{zhang2022data}, the authors proposed to use data augmentation techniques to enhance the robustness of the neural precoding methods for designing beamforming in the MISO system. However, their methods rely on adding noise to the input CSI, which may not capture the true diversity and complexity of the new deployment domain.
A shortcoming of many existing studies in transfer learning for wireless communications is that both the training and deployment scenarios are simulated using statistical channel models, which may not capture the full extent of the domain shift.

In this paper, we focus on enhancing the adaptability and generalization capabilities of DL-based \gls{HBF} in diverse channel conditions while addressing the resource-intensive nature of fine-tuning. Our goal is to seamlessly integrate DL-based HBF into the dynamic realm of mMIMO systems. Inspired by the success of meta-learning and domain generalization, we aim for swift adaptation. Additionally, we introduce an innovative data augmentation technique in unsupervised fine-tuning to reduce the resource demands for dataset gathering. Based on our evaluations, we demonstrate our method's swift adaptation, requiring fewer fine-tuning samples, and its superior performance over existing DL-based solutions. Notably, it can achieve good performance even in scenarios with changes in antenna configurations. Finally, the experimental validation provides insight into the transition from statistical channel models to real-world scenarios. 
\section{System model} \label{Sec:Baseline}
Let us consider a multi-user \gls{mMIMO} system consisting of a \gls{BS} equipped with $N_{\sf{T}}$ antennas and $N_{\sf{RF}}$ \gls{RF} chains serving $N_{\sf{U}}$ single antenna users simultaneously. For the downlink transmission, \gls{HBF} precoders are employed by the \gls{BS}.
The signal received by each user can be written as
\begin{equation} \label{recv_sig}
\mb{y}_u =  \mb{h}_{u}^{\dagger} \mb{A} \sum_{\forall u}  \mb{w}_{u} x_u + \bs{\eta},
\end{equation}
where $\mb{h}_{u} \in \mathbb{C}^{N_{\sf{T}} \times 1}$ stands for the channel vector from the user index $u$ to the $N_{\sf{T}}$ antennas at the \gls{BS}, $\mb{x} = [x_1, ..., x_u, ..., x_{N_{\sf{U}}}]$ is the vectors of transmitted symbol for all users, normalized to $\mathop{\mathbb{E}}[\mb{x}\mb{x}^{\dagger}] = \frac{1}{N_{\sf{U}}} \mathbbm{1}_{N_{\sf{U}}}$, and $\bs{\eta}$ is the \gls{AWGN} term with noise power $\sigma^2$. The \gls{HBF} vectors consist of a digital baseband precoder $\mb{W} = [\mb{w}_{1}, ..., \mb{w}_{u}, ..., \mb{w}_{N_{\sf{U}}}] \in \mathbb{C}^{N_{\sf{RF}} \times N_{\sf{U}}}$ and \gls{AP} which is defined by a $N_{\sf{T}}\times N_{\sf{RF}}$ matrix $\mb{A} \in \mathbb{C}^{N_{\sf{T}} \times N_{\sf{RF}}}$, where $[\mb{A}]_{n,m} = e^{j\varphi_{n,m}}$ is the state of the \gls{PS} connected between the $n^{\text{th}}$ antenna and $m^{\text{th}}$ RF chain by the phase-shift of $\varphi_{n,m}$. The \gls{SINR} of the $u^{\text{th}}$ user is given by
\begin{align}
    \label{eq:SINR_HBF}
    \text{SINR}(\mb{A},\mb{w}_{u}) & = \frac{ \big|\mb{h}^{\dagger}_{u} \mb{A} \mb{w}_{u} \big|^2}{\sum_{j \neq u} \big|\mb{h}^{\dagger}_{u}\mb{A} \mb{w}_{j} \big|^2 + \sigma^2} \, .
\end{align}
For a given hybrid beamformer $(\mb{A},\mb{W})$, the sum-rate is
\begin{align}
\label{eq:sumRate_HBF}
    R(\mb{A},\mb{W}) = \sum_{\forall u} \text{log}_2 \Bigl(  1+ \text{SINR}(\mb{A},\mb{w}_{u}) \Bigr) \,.
\end{align}
Then, the \gls{HBF} design consists of finding the precoder matrices $\mb{W}$ and $\mb{A}$ that maximize the sum-rate in \eqref{eq:sumRate_HBF} subject to a maximum transmission power $P_{\sf{max}}$. More formally, in training backbone model and deployment scenario, we seek to solve the following optimization problem:
\begin{subequations}  \label{SEE-max-prb}
\begin{eqnarray} 
&\underset{\mb{A},\mb{W}}{\max} &  R(\mb{A},\mb{W}) = \underset{\mb{A},\mb{w}_u}{\max} \sum_{\forall u}  \text{log}_2 \Bigl(  1+ \text{SINR}(\mb{A},\mb{w}_{u}) \Bigr), \\
& \text{s.t.} & \sum_{\forall u} \mb{w}_{u}^{\dagger} \mb{A}^{\dagger}  \mb{A} \mb{w}_{u} \leq P_{\sf{max}}. \label{cnt2}
\end{eqnarray}
\end{subequations}%

\subsection{Problem Definition}
In this paper, we study how effectively the DNN, initially trained in a statistical channel environment, can adapt its HBF strategies to the dynamic and unpredictable nature of practical wireless communication scenarios. 
Training with a statistical channel model offers a practical advantage. By utilizing this simplified yet representative model, we can train a robust backbone model in a cloud-based environment with a large available statistical dataset, detaching the training process from the intricacies of specific channel conditions. This decoupling allows us to create a flexible and generalized \gls{DNN} in varying channels that is applicable across numerous \gls{BS}.

The statistical channel model utilized in this study offers a practical and computationally efficient tool for training our DNN for HBF in mMIMO systems, with a specific focus on Line-of-Sight (LOS) scenario similar to the deployment phase. To generate statistical channel models for the backbone training, $\mb{h}_{u}$ is expressed as \cite{mukherjee2022performance}
\begin{equation}
\label{eq:5}
\mb{h}_{u}=  \mathbf{a}(\phi, \theta,\mathbf{n}) \sqrt{G_{BS} G_u}\left(\frac{\ell_{BS} \ell_u }{4 \pi x_{u}^{\gamma}}\right) e^{j 2 \pi \frac{x_{u}}{\lambda_c}} \, ,
\end{equation}
where $(G_{BS}, \ell_{BS})$ and $(G_u, \ell_u)$ are the gains and the height corresponding with the antennas of the BS and $u$th UE, respectively. The three-dimensional distance between the BS and $u$th UE is denoted by $x_u$ and  $\lambda_c$ represents the carrier wavelength. The exponent path loss factor ($\gamma$) is a parameter that describes how the signal power decreases with distance in a wireless communication system.
In addition, $\mathbf{a}(\phi, \theta, \mathbf{n})$ denotes the antenna array response vector of the \gls{BS}, where $\mathbf{n}$ is the base station antenna array dimensions in each direction $\mathbf{n}=[N_x, N_y, N_z]$, and $\phi, \theta$ stand for azimuth and elevation angles of arrival at the BS, respectively. The general array response vector $\mathbf{a}(\phi, \theta, \mathbf{n})$ is decomposed as
\begin{equation}
\begin{aligned}
\mathbf{a}(\phi, \theta, \mathbf{n}) = &
~\bs{\omega}\left(\psi_z(\phi, \theta), N_z\right) \otimes \\ &~\bs{\omega}\left(\psi_y(\phi, \theta), N_y\right) \otimes \\ &~\bs{\omega}\left(\psi_x(\phi, \theta), N_x\right),
\end{aligned}
\end{equation}
with the components given by
\begin{equation}
\bs{\omega}\left(\psi, N\right) = \left[ 1, ~\mathrm{e}^{j  2 \pi \frac{d_{BS}}{\lambda_c} \psi}, ~\ldots, ~\mathrm{e}^{j 2 \pi \frac{d_{BS}}{\lambda_c} \left(N-1\right) \psi} \right]^T,    
\end{equation}
where $d_{BS}$ is the spacing between antennas at the BS, and for each direction, $\psi$ is expressed as
\begin{equation}
\begin{aligned}
\psi_x(\phi, \theta) &= \sin(\theta) \cos(\phi)\\ 
\psi_y(\phi, \theta) &=  \sin(\theta) \sin(\phi)\\ 
\psi_z(\phi, \theta) &=  \cos(\theta). \\ 
\end{aligned}
\end{equation}

For the testing and deployment phase, we transition to a more complex and realistic channel model, namely the ray-tracing channel model. We used deepMIMO channel model \cite{alkhateeb2019deepmimo} for our deployment scenario. Unlike the statistical model, the ray-tracing model provides a high-fidelity representation of the wireless channel by simulating the propagation of electromagnetic waves in a detailed 3D environment. This approach considers various factors, including reflections, diffractions, and scattering, to capture the intricacies of real-world propagation conditions accurately. By training the DNN with a statistical channel model and subjecting the DNN-based HBF model to the complexities of the ray-tracing realistic channel model, we aim to uncover crucial insights into its adaptability and robustness in real-world conditions.

One of the main research challenges of our study is to develop a robust backbone model characterized by its adaptability to variations in the environment and \gls{BS} configurations. In particular, this study emphasizes the importance of three critical parameters in mMIMO systems: the antenna array response vector, the height of the base station, and the path loss exponent factor. The antenna array response vector is a critical component that can vary based on different antenna heights and possess different numbers of elements in each direction. For instance, as illustrated in Table~\ref{tab:my-table}, the backbone model was initially trained under different antenna array settings and subsequently deployed in diverse configurations. In this context, it is important to note that we assume $N_{\sf{T}} = 64$, and the subscript $i$ within $\mb{n}_{(i)}$ indicates the directions of the antenna array. For instance, in the case of $\mb{n}_{yz}$, it implies the deployment of 64 antennas along the $y$ and $z$ axes with respective quantities of $\mathbf{n}_{yz}=[1, 8, 8]$. The outcomes reveal a significant degradation in zero-shot\footnote{i.e., without fine-tuning} performance when the model was deployed in configurations for which it was not originally tailored (as observed in the non-diagonal elements of the tables). Therefore, it becomes apparent that the alteration of the antenna array response vector can lead to a substantial decline in model performance.

The height of the base station is also another critical parameter in mMIMO systems, influencing signal propagation and coverage. To underscore its significance, we conducted experiments highlighting the impact of varying base station heights during training and deployment. For example, training the backbone model for a base station with a height of 12m and deploying it in the ray-tracing scenario with a different height of 6m resulted in a significant drop of $43.24\%$ in zero-shot performance compared to near-optimal~\cite{hojatian2022flexible}. This highlights the need to consider base station height variations during training for improved model robustness and zero-shot performance in diverse deployment scenarios.

The path loss exponent factor captures the power falloff of the signal relative to the distance from the transmitter and depends on the carrier frequency, environment, obstructions, etc. It is an important parameter for modeling the channel behavior and performance of mMIMO systems in different scenarios. The path loss exponent factor is used to increase the generalization of the channel model to unknown environments. By varying this parameter, different levels of path loss can be simulated and analyzed. This allows us to evaluate the robustness and adaptability of mMIMO systems under various propagation conditions.

\begin{table}[t]
\centering
\caption{Impact of Different Antenna Configuration \\on Zero-Shot Sum Rate (b/s/Hz) ($-\log\sigma^2=13$)}
\vspace{-8pt}
\label{tab:my-table}
\resizebox{\columnwidth}{!}{%
\begin{tabular}{lcccccccc}
\cline{3-9}
\multicolumn{1}{c}{} & 
\multicolumn{1}{c|}{} & 
\multicolumn{7}{c|}{Deployment} \\ \cline{3-9} 
\multicolumn{1}{c}{} & 
\multicolumn{1}{c|}{} & 
$\mathbf{n}_{x}$ & 
$\mathbf{n}_{y}$ & 
$\mathbf{n}_{z}$ & 
$\mathbf{n}_{xy}$ &
$\mathbf{n}_{xz}$ &
$\mathbf{n}_{yz}$ & 
\multicolumn{1}{c|}{$\mathbf{n}_{xyz}$} \\ \cline{1-2}
\hline
\multicolumn{1}{|l|}{} & 
\multicolumn{1}{c|}{$\mathbf{n}_{x}$} &
\cellcolor[HTML]{FBCDEB}15.89 &
\cellcolor[HTML]{9698ED}6.53 &
\cellcolor[HTML]{9698ED}1.54 & 
\cellcolor[HTML]{FBCDEB}10.42 & 
\cellcolor[HTML]{CBCEFB}8.23 & 
\cellcolor[HTML]{9698ED}6.34 & 
\multicolumn{1}{c|}{\cellcolor[HTML]{CBCEFB}7.34} \\
\multicolumn{1}{|l|}{} & 
\multicolumn{1}{c|}{$\mathbf{n}_{y}$} &
\cellcolor[HTML]{CBCEFB}8.92 &
\cellcolor[HTML]{FBCDEB}15.39 &
\cellcolor[HTML]{9698ED}1.57 &
\cellcolor[HTML]{CBCEFB}8.33 &
\cellcolor[HTML]{CBCEFB}7.63 &
\cellcolor[HTML]{CBCEFB}7.83 & 
\multicolumn{1}{c|}{\cellcolor[HTML]{9698ED}6.51} \\
\multicolumn{1}{|l|}{} & 
\multicolumn{1}{c|}{$\mathbf{n}_{z}$} &
\cellcolor[HTML]{9698ED}1.23 &
\cellcolor[HTML]{9698ED}1.36 &
\cellcolor[HTML]{FBCDEB}14.45 &
\cellcolor[HTML]{9698ED}1.54 &
\cellcolor[HTML]{9698ED}5.84 &
\cellcolor[HTML]{9698ED}4.53 & 
\multicolumn{1}{c|}{\cellcolor[HTML]{9698ED}6.34} \\
\multicolumn{1}{|l|}{} & 
\multicolumn{1}{c|}{$\mathbf{n}_{xy}$} &
\cellcolor[HTML]{FBCDEB}11.13 & 
\cellcolor[HTML]{CBCEFB}8.49 & 
\cellcolor[HTML]{9698ED}1.26 & 
\cellcolor[HTML]{FBCDEB}14.67 & 
\cellcolor[HTML]{9698ED}4.64 & 
\cellcolor[HTML]{9698ED}5.69 & 
\multicolumn{1}{c|}{\cellcolor[HTML]{9698ED}3.18} \\
\multicolumn{1}{|l|}{} & 
\multicolumn{1}{c|}{$\mathbf{n}_{xz}$} &
\cellcolor[HTML]{CBCEFB}8.01 & 
\cellcolor[HTML]{CBCEFB}7.51 & 
\cellcolor[HTML]{9698ED}5.64 &
\cellcolor[HTML]{9698ED}4.15 &
\cellcolor[HTML]{FBCDEB}13.86 &
\cellcolor[HTML]{9698ED}5.51 & 
\multicolumn{1}{c|}{\cellcolor[HTML]{CBCEFB}7.61} \\
\multicolumn{1}{|l|}{} &
\multicolumn{1}{c|}{$\mathbf{n}_{yz}$} &
\cellcolor[HTML]{9698ED}6.21 & 
\cellcolor[HTML]{CBCEFB}8.14 & 
\cellcolor[HTML]{9698ED}4.16 & 
\cellcolor[HTML]{9698ED}5.77 & 
\cellcolor[HTML]{9698ED}6.36 & 
\cellcolor[HTML]{FBCDEB}14.24 &
\multicolumn{1}{c|}{\cellcolor[HTML]{9698ED}5.76} \\
\multicolumn{1}{|l|}{\multirow{-7}{*}{\rotatebox[origin=c]{90}{Backbone}}} & 
\multicolumn{1}{c|}{$\mathbf{n}_{xyz}$} &
\cellcolor[HTML]{CBCEFB}7.69 &
\cellcolor[HTML]{9698ED}5.41 & 
\cellcolor[HTML]{9698ED}5.26 &
\cellcolor[HTML]{9698ED}5.36 & 
\cellcolor[HTML]{CBCEFB}8.13 & 
\cellcolor[HTML]{9698ED}5.32 & 
\multicolumn{1}{c|}{\cellcolor[HTML]{FBCDEB}12.64} \\ \hline
\multicolumn{2}{|c|}{near-optimal \cite{hojatian2022flexible}} &
\multicolumn{1}{l}{22.36} &
\multicolumn{1}{l}{21.26} &
\multicolumn{1}{l}{18.64} &
\multicolumn{1}{l}{20.91} &
\multicolumn{1}{l}{19.18} &
\multicolumn{1}{l}{18.54} 
& \multicolumn{1}{l|}{18.37} \\ \hline
\end{tabular}%
}
\vspace{-10pt}
\end{table}

\vspace{-10pt}
\section{Sage-HB-Net} 
\label{Sec:Proposed}
This section presents our proposed adaptive deep learning method for mMIMO HBF, 
which includes a data augmentation method in the deployment phase that allows fine-tuning of the model with a small dataset.

\subsection{Backbone Training}
Domain generalization, also known as out-of-distribution generalization, is a technique applied in scenarios where data availability in the deployment environment is restricted. This limitation may arise due to privacy considerations or the necessity for real-time learning~\cite{shen2021towards}. 
On the other hand, meta-learning, often described as ``learning to learn,'' introduces a hierarchical structure in AI. This approach empowers AI systems to learn rapidly, adapt to changing environments, and apply their knowledge to a broader array of tasks. Instead of relying on copious data from similar tasks encountered in the past to tackle a new challenge, meta-learning focuses on the capacity to acquire the skill of learning itself.

In our work, one objective is to significantly reduce the data dependency of our model during the deployment phase, while concurrently enhancing its energy and spectral efficiency. To train such a model, we use the \gls{MLDG} method introduced in \cite{li2018learning}. \gls{MLDG} employs a training technique with the goal of improving the adaptability of a model to unseen domains. Within the meta-training, the aim of the model is to reduce the loss within a specific set of domains, referred to as the training domains. Simultaneously, it seeks to ensure that the optimization gradients selected also lead to enhancements in a different subset of domains known as the testing domains. Leveraging the \gls{MLDG} algorithm as our foundation, we intend to formulate a meta-learning and domain generalization algorithm tailored to adaptive \gls{mMIMO} \gls{HBF}. We aim to achieve a superior adaptation and higher level of zero-shot performance compared with other existing adaptive methods. 

The proposed method is outlined in Algorithm~\ref{Alg:1}, which we now explain in detail.
In the first stage, known as ``Backbone-Training,'' the algorithm is initialized with model parameters ($\Theta$) and learning rates for each updating part ($\alpha, \beta, \epsilon$). It operates on a set of statistical channel model domains ($\bs{\mathcal{D}}$) with the following key steps. The algorithm initiates by randomly dividing available domains into two subsets: $\bs{\mathcal{D}}^{train}$ for training and $\bs{\mathcal{D}}^{gen}$ for generalization, ensuring adaptability to new, previously unseen environments. In the meta-training phase, for each domain within the training set ($\hat{\mathcal{D}} \in \bs{\mathcal{D}}^{train}$), the algorithm computes gradients of the loss function with respect to the model parameters ($\nabla_{\Theta}\mathcal{L}(\hat{D};\Theta)$). These gradients represent the learning progress for each domain, which are accumulated in $L$. Following the accumulation of gradients, the algorithm updates the model parameters using $L$ to enhance its performance within the training domains ($\Theta^{\prime} = \Theta - \alpha L$). This step allows the model to better adapt to the characteristics of the known training environments. In the meta-testing phase, similar to meta-training, the algorithm computes gradients ($\nabla_{\Theta^{\prime}}\mathcal{L}(\hat{\mathcal{D}};\Theta^{\prime})$) for the domains in the generalization set ($\hat{\mathcal{D}} \in \mathcal{D}^{gen}$). These gradients are accumulated in $L^{\prime}$ and are used to evaluate the model's ability to generalize to new, unseen environments. Subsequently, the algorithm performs a meta-update, adjusting the model parameters based on the aggregated gradients from both the training and generalization domains ($\Theta \leftarrow \Theta - \lambda(L + \beta L^{\prime})$). This meta-update is crucial for enhancing the model's generalization capabilities, making it more adaptable to different environments. The main objective is to reduce losses in both domains simultaneously in a synchronized way.

\begin{algorithm}[t]
    \label{Alg:1}
  \caption{SAGE-HB-Net}
  \vspace{2pt}
  \centerline{ \textbf{Backbone-Training}}
  \vspace{2pt}
  \hrule
  \vspace{2pt}
  \KwIn{Statistical channel model domains $\bs{\mathcal{D}}$}
  \SetKwInOut{Init}{Initialize}
  \SetKw{IN}{in}
  \SetKw{each}{each}
  \newcommand{\forcond}{$e=1$ \KwTo $Epochs$}
  \newcommand{\forcondd}{$e=1$ \KwTo $Epochs$}
  \newcommand{\forcondom}{$i=1$ \KwTo $m$}
  \newcommand{\forcondou}{$j=1$ \KwTo $N_U$}
  \newcommand{\fordomains}{\each $\hat{D}$ \IN $\bs{\mathcal{D}}^{train}$ }
  \newcommand{\fordomainsgen}{\each $\hat{D}$ \IN $\bs{\mathcal{D}}^{gen}$ }
  \Init{Model parameters $\Theta$, Learning rates $\alpha, \beta, \epsilon$}  
  
 \For{\forcond}{
  \textbf{Divide} $\bs{\mathcal{D}} \rightarrow \bs{\mathcal{D}}^{train}$ and $\bs{\mathcal{D}}^{gen}$~~~~~~~~~~~~~~~\tikzmark{right}\\
  $L \gets 0$\tikzmark{top}\\
  \For{\fordomains}{
  $L \leftarrow L + \frac{\nabla_{\Theta}\mathcal{L}(\hat{D};\Theta)}{|\bs{\mathcal{D}}^{train}|}$\\
    }
  \textbf{Update} $\Theta^{\prime} = \Theta - \alpha L$ \tikzmark{bottom}\\
  \AddNote{top}{bottom}{right}{Meta-Training}
  $L^{\prime} \gets 0$\tikzmark{top}\\
  \For{\fordomainsgen}{
  $L^{\prime} \leftarrow L^{\prime} + \frac{\nabla_{\Theta^{\prime}}\mathcal{L}(\hat{D};\Theta^{\prime})}{{|\bs{\mathcal{D}}^{gen}|}}$\tikzmark{bottom}\\
    }
    \AddNote{top}{bottom}{right}{Meta-Tesing}
    \textbf{Meta-update:} $\Theta \leftarrow \Theta -  \epsilon(L + \beta L^{\prime})$\\
    }
  \hrule
  \vspace{2pt}
  \centerline{\textbf{Fine-Tuning}}
  \vspace{2pt}
  \hrule
  \vspace{2pt}
  \KwIn{Flattened dataset $\mathbf{F}$, 
  Augmentation size $m$}
  \Init{Model parameters $\Phi \leftarrow \Theta$, Learning rate $\tau$}
  \vspace{2pt}
  $D^{Aug} \gets \emptyset$\tikzmark{top}\\
  \For{\forcondom}{
    $S \gets$ all-zero matrix of size $N_T \times N_U$ \\
    \For{\forcondou}{
    $k \sim \mathcal{U}(1, |\mathbf{F}|)$\\
    $S[:,j]=F[:,k]$\\
  }
  $D^{Aug} \leftarrow D^{Aug} \cup \{S\}$\tikzmark{bottom}\\
  }\AddNote{top}{bottom}{right}{Data Augmentation}
  \For{\forcondd}{
  $\Phi \leftarrow \Phi - \tau \nabla_{\Phi}\mathcal{L}(\mathcal{D}^{Aug};\Phi)$
  }
  \end{algorithm}

\subsection{Fine Tuning}
\begin{figure}[t]
    \vspace{-10pt}
    \centering
    \includegraphics[width=.8\columnwidth]{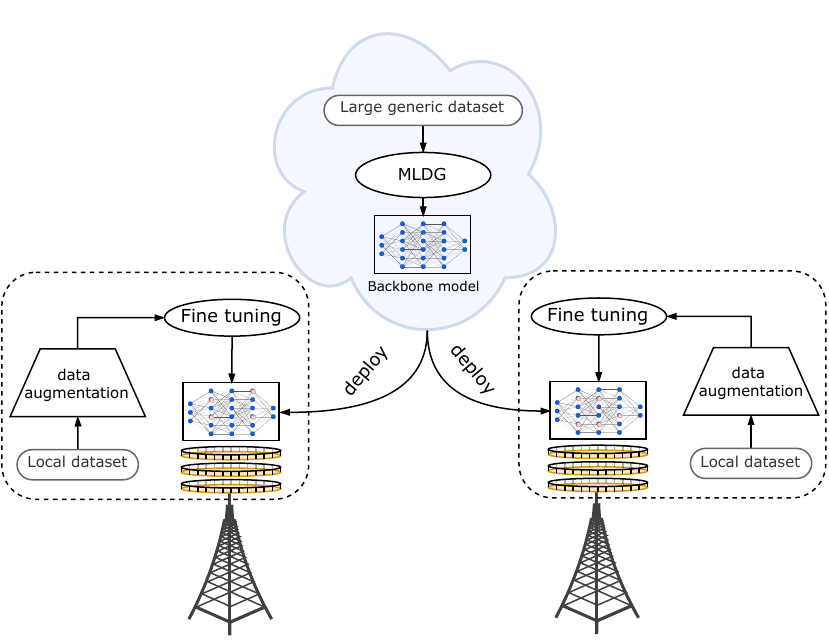}
    \caption{SAGE-HB-Net Backbone Training Procedure and Deployment Across Diverse Base Stations}
    \vspace{-5pt}
    \label{fig:sys_fig}
\end{figure}

In the second stage, known as ``Fine-Tuning,'' the BS is deployed in a specific new environment and it starts the unsupervised fine-tuning.
Firstly, the BS initializes a new set of model parameters~($\Phi$) with the parameters downloaded from the backbone model~($\Theta$) in the cloud, as illustrated in Fig~\ref{fig:sys_fig}. Next, the Data Augmentation step is performed. 
Let $\mathcal{D}$ be a dataset, where each element with index $d$ corresponds to a matrix $\mathbf{H}^{(d)}$ of size $N_T \times N_U$, and in turn each column of $\mathbf{H}^{(d)}$ corresponds to a CSI vector $\mathbf{h}_u^{(d)}$ for user $u$. 
The flattened dataset $\mathbf{F}$ is defined as an $N_T \times N_U |\mathcal{D}|$ matrix obtained by stacking the columns of all the matrix elements $\mathbf{h}^{(i)}$ of $\mathcal{D}$, that is
\vspace{-5pt}
\[
\mathbf{F} = \begin{bmatrix} \mathbf{h}_{1,:}^{(1)} & \mathbf{h}_{2,:}^{(1)} & \ldots & \mathbf{h}_{N_U,:}^{(1)} & \mathbf{h}_{1,:}^{(2)} & \ldots & \mathbf{h}_{N_U,:}^{(|\mathcal{D}|)} \end{bmatrix} \, .
\]
The BS randomly selects subsets of users' CSI ($S$) from $\mathbf{F}$, effectively creating a new dataset ($D^{Aug}$). Since user positions are independent and do not affect one another's channels, by combining different channels of individual users, we are able to quickly grow the multi-user dataset and cover a wider range of possible scenarios. The maximum value of the data augmentation size is $m=\binom{|\mathcal{D}| N_U}{N_U}$. 
Finally, in the parameter update step, the model parameters ($\Phi$) in the BS are updated using stochastic gradient descent with unsupervised learning. The gradients of the loss function with respect to the augmented data ($\nabla_{\Phi}\mathcal{L}(\mathcal{D}^{Aug};\Phi)$) guide the parameter updates. This fine-tuning process ensures that the model is optimized and ready to perform effectively in the new deployment environment.

\subsection{SAGE-HB-Net Architecture}
\begin{figure}[t]
    \vspace{-5pt}
    \centering
    \includegraphics[width=\columnwidth]{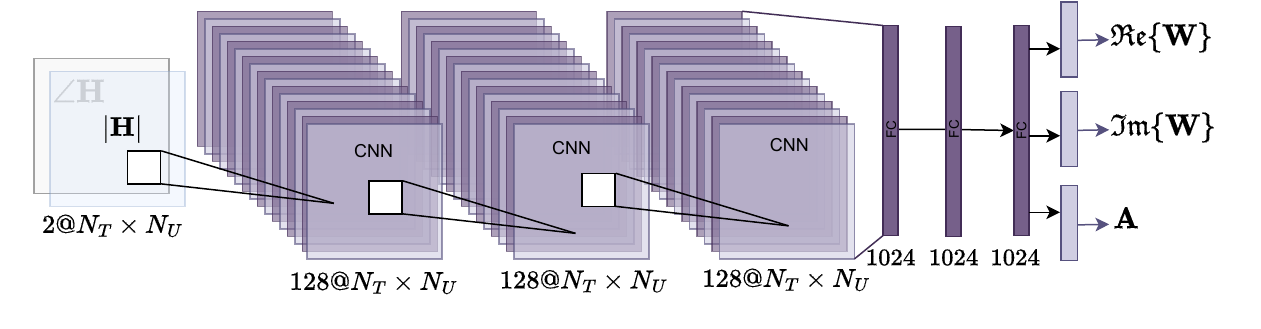}
    \vspace{-20pt}
    \caption{Proposed DNN architecture for Hybrid Beamforming}
    \vspace{-15pt}
    \label{fig:arch}
\end{figure}

Our DNN architecture for hybrid beamforming is illustrated in Figure \ref{fig:arch}. This design is customized to the specific requirements of our problem. Unlike previous unsupervised learning models for hybrid beamforming that use the real and imaginary parts of channel data as inputs~\cite{hojatian2021unsupervised}, we find that utilizing the amplitude and phase of the channels as the DNN inputs yields superior performance, particularly when dealing with varying antenna configurations.
The architecture comprises three consecutive CNN layers with $128$ channels, followed by three fully connected layers (FC) with $1024$ neurons. These components collectively enable the model to effectively capture complex channel interactions and patterns. Subsequently, we extract the analog precoder and also real and imaginary parts of the digital precoder as output. The architecture incorporates ReLU activation functions, 50\% dropout layers to prevent overfitting, and batch normalization for enhanced convergence and stability in adapting to mMIMO environments.

\section{Transfer Learning Numerical Results} \label{Sec:Simulation}
In order to benchmark as accurately as possible the adaptability of the DNN models, we model the deployment scenario using a ray-tracing environment, while only statistical models are allowed to be used for backbone training.
The performance of SAGE-HB-Net and the baseline solutions was numerically evaluated using the \textsc{PyTorch} \gls{DL} framework. 
In the process of training a backbone model, a batch size of $1000$ was employed to process data efficiently. During backbone training, the learning rate parameters are set to $\alpha=10^{-4}$, $\epsilon=10^{-5}$, and 
$\beta=1$. We use $\tau=10^{-3}$ as a fine-tuning learning rate. 

\subsection{Backbone Model Training}
Our backbone training dataset is constructed using statistical channel models. To ensure the model's adaptability and robustness, we introduce diversity in our training data by considering various source domains. In practical base station scenarios, the antenna configuration may vary widely due to spatial constraints, user distribution, or hardware limitations. 
In this phase, the selection of specific antenna configurations serves a strategic purpose aimed at benchmarking the adaptability of the backbone model. We create a training set that encourages the model to generalize effectively across diverse antenna array responses, but we specifically omit configurations that will be tested at deployment time and, in particular, avoid any two-dimensional configurations. Excluding certain configurations ensures that the model encounters scenarios it has not seen before. The chosen configurations are $\textbf{n} \in \{[1,1,64], [1,64,1], [64,1,1], [4,4,4]\}$.

We further expand our source domains by introducing different path loss exponent values to enhance the model's ability to capture the nuances of signal attenuation and propagation in various environments, mirroring real-world scenarios. Specifically, we consider four path loss exponents $\gamma \in \{1.3, 1.5, 1.7, 2.0\}$. Thus, in total, the source domain set contains 16 domains.

In addition to the diverse antenna array responses and path loss exponents, our statistical channel models incorporate integer-varying base station heights. Specifically, the parameter $\ell_{BS}$ in equation (\ref{eq:5}) is a random variable uniformly sampled once from the set $\{6, 7,\dots, 12 \}$ for each element of the dataset, ensuring variability in base station height and contributing to the model's exposure to a broad range of scenarios for enhanced adaptability and robust performance in mMIMO environments with different base station configurations.
\vspace{-5pt}
\subsection{Deployment Scenario}
Scenario ``O1-$28$~GHz'' of the deepMIMO channel model~\cite{alkhateeb2019deepmimo} was used in the generation of a dataset for the evaluation. This scenario involves placing several users' locations randomly in streets bounded by constructions. It involves the third \gls{BS} with height of 6m and equipped with $N_{\sf{T}} = 64$ antennas and $N_{\sf{RF}} = 8$ RF chains serving $N_{\sf{U}}=4$ users randomly located in a dedicated area~(in deepMIMO channel model $\texttt{active\_user\_first} = 400$ and $\texttt{active\_user\_last} = 1400$). This unseen scenario provides a realistic environment for assessing the adaptability and performance of the proposed SAGE-HB-Net algorithm in a mMIMO system. Note that all methods were optimized independently, and we applied the data augmentation method across all methods. This ensures fairness and comparability in our study. In addition, all DL-based methods were evaluated on the same network architecture. The training sets used for backbone training and fine-tuning were the same across all DL-based methods.

\begin{table}[t]
\centering
\caption{{\footnotesize Impact of Antenna Configuration Variability on SAGE-HB-Net's Zero Shot sum rate (b/s/Hz), Noise power: $-\log\sigma^2=13$}}
\label{tab:my-table2}
\resizebox{\columnwidth}{!}{%
\begin{tabular}{lcccccccc}
\cline{3-9}
\multicolumn{1}{c}{} & \multicolumn{1}{c|}{} & \multicolumn{7}{c|}{Deployment} \\ \cline{3-9} 
\multicolumn{1}{c}{} &
\multicolumn{1}{c|}{} & 
$\mathbf{n}_{x}$ & 
$\mathbf{n}_{y}$ & 
$\mathbf{n}_{z}$ & 
$\mathbf{n}_{xy}$ & 
$\mathbf{n}_{xz}$ & 
$\mathbf{n}_{yz}$ & 
\multicolumn{1}{c|}{$\mathbf{n}_{xyz}$} \\ \cline{1-2}
\hline
\multicolumn{2}{|c|}{SAGE-HB-Net} &
\cellcolor[HTML]{FBCDEB}11.63 &
\cellcolor[HTML]{FBCDEB}10.82 & 
\cellcolor[HTML]{FBCDEB}11.24 & 
\cellcolor[HTML]{CBCEFB}9.62 & 
\cellcolor[HTML]{CBCEFB}9.45 & 
\cellcolor[HTML]{CBCEFB}9.79 & 
\multicolumn{1}{c|}{\cellcolor[HTML]{FBCDEB}10.36}
\\ \hline
\multicolumn{2}{|c|}{near-optimal \cite{hojatian2022flexible}} &
\multicolumn{1}{l}{22.36} &
\multicolumn{1}{l}{21.26} &
\multicolumn{1}{l}{18.64} &
\multicolumn{1}{l}{20.91} &
\multicolumn{1}{l}{19.18} &
\multicolumn{1}{l}{18.54} &
\multicolumn{1}{l|}{18.37} \\ \hline
\end{tabular}%
}
\vspace{-15pt}
\end{table}
Table~\ref{tab:my-table2} reports a significant improvement in zero-shot performance for SAGE-HB-Net compared to the non-generalized models discussed in Table~\ref{tab:my-table}. In Table~\ref{tab:my-table}, the model exhibited a substantial degradation in zero-shot sum rate, indicating challenges in adapting to diverse antenna array configurations. In contrast, Table~\ref{tab:my-table2} highlights the enhanced adaptability of SAGE-HB-Net, consistently achieving high sum rates across various deployment scenarios with varying antenna array response vectors. Although SAGE-HB-Net achieves approximately half of the near-optimal sum rate for zero-shot performance, its ability to maintain robust performance in diverse settings underscores its effectiveness in practical mMIMO deployments. This improvement emphasizes the ability of SAGE-HB-Net to address challenges related to antenna configuration variability, showcasing its potential for real-world applications.

\begin{figure}[t]\centering\includegraphics[width=\columnwidth]{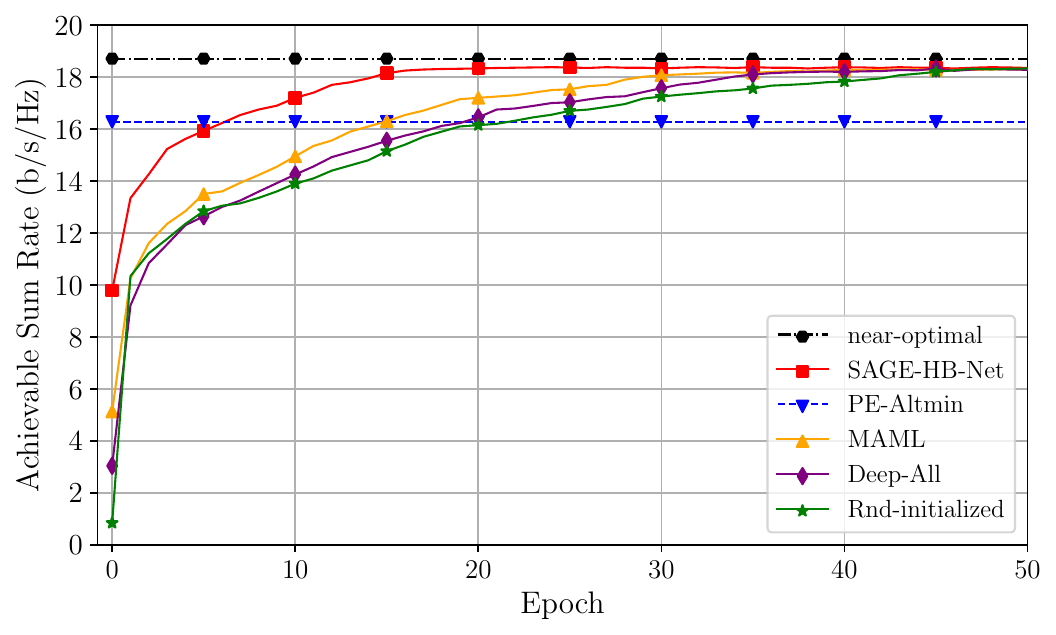}
    \vspace{-15pt}
    \caption{Fine-tuning on DeepMIMO with $\textbf{n}=[1,8,8]$, $|\mathcal{D}|=10^{6}$, and $-\log\sigma^2=13$.}
    \vspace{-20pt}
    \label{fig:FT}
\end{figure}

In Figure \ref{fig:FT}, we present an evaluation of the adaptability of our proposed solution in comparison to other state-of-the-art adaptive methods. The near-optimal performance baseline is evaluated by training the DNN of \cite{hojatian2022flexible} on DeepMIMO for $200$ epochs. 
Additionally, PE-AltMin \cite{yu2016alternating} is included in the comparison to provide an example of the performance achievable with conventional optimization-based methods. Adaptive methods include \gls{MAML}~\cite{yuan2020transfer}, and Deep-All, which is a training approach in which we aggregate datasets from all source domains to build the backbone model. An additional benchmark, ``Rnd-initialized,'' assesses adaptability to new deployment scenarios without pre-trained knowledge.

The results show that SAGE-HB-Net achieves substantially higher zero-shot sum-rates, surpassing the sum-rates of \gls{MAML}, Deep-All, and Rnd-initialized approaches by factors of approximately $1.9$, $3.2$, and $11.6$ times, respectively. Furthermore, the convergence speed of the proposed solution stands out as another notable advantage. We see that in just $15$ epochs, SAGE-HB-Net approaches the near-optimal performance. In contrast, \gls{MAML} requires around 30 epochs, Deep-All approximately $35$ epochs, and Rnd-initialized about $45$ epochs to reach the same sum-rate performance. In unseen environments, SAGE-HB-Net shows enhanced adaptability and efficiency in harnessing the potential of DL-based beamforming.

\begin{figure}[t]
    \centering\includegraphics[width=0.95\columnwidth]{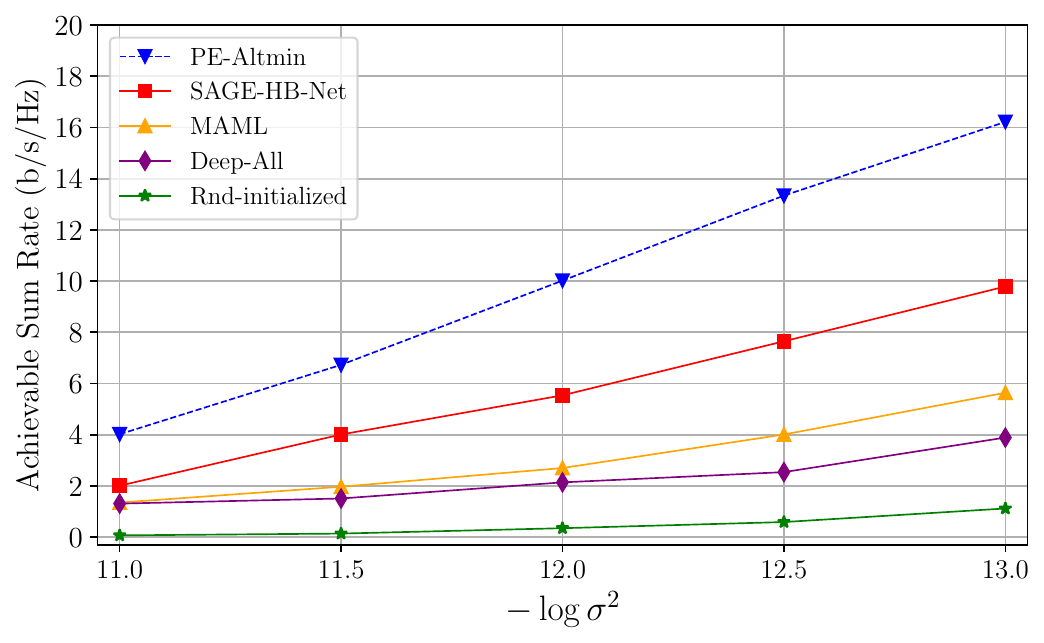}
    \vspace{-10pt}
    \caption{ Zero Shot performance for different noise powers on DeepMIMO, with $\textbf{n}=[1,8,8]$.}
    \vspace{-15pt}
    \label{fig:SR-ZS}
\end{figure}

We assess the zero-shot sum-rate performance of our proposed method across various noise power values ($\sigma^2$), as depicted in Figure \ref{fig:SR-ZS}. 
Our results indicate that our proposed solution outperforms other DL-based approaches in terms of zero-shot sum-rate performance. It is important to note that while the initial sum-rate of our method may be lower than that of PE-AltMin, this observation is specific to the zero-shot sum-rate scenario. As shown in Figure \ref{fig:FT}, after a few fine-tuning iterations, the sum-rate result of all DL-based methods surpasses that of PE-AltMin.

\begin{figure}[t]
    \centering\includegraphics[width=0.95\columnwidth]{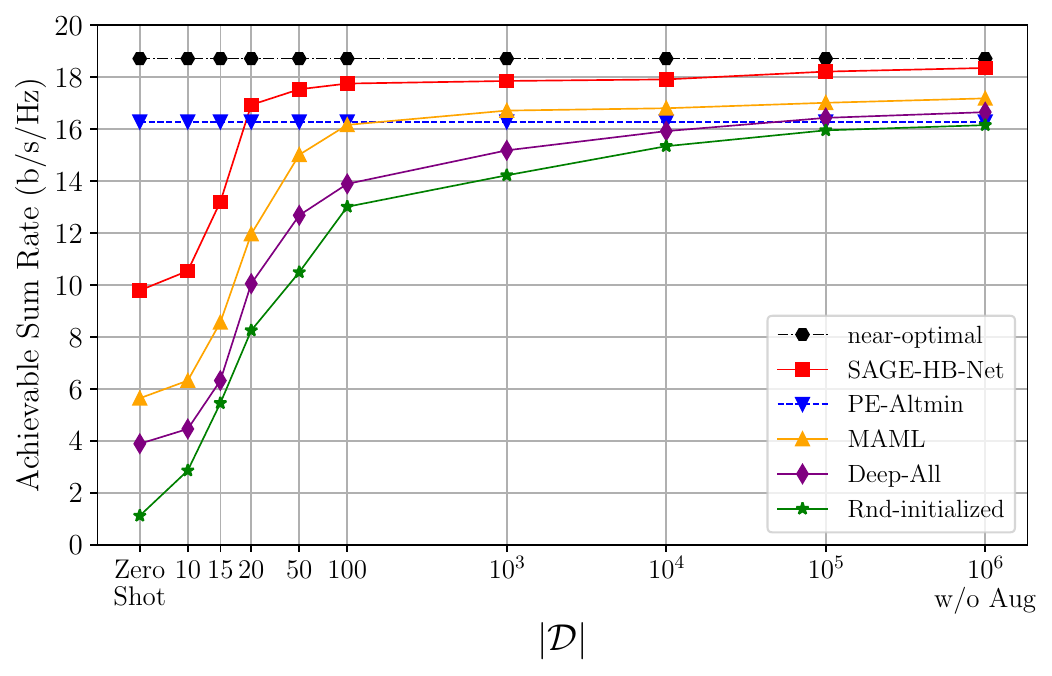}
    \vspace{-15pt}
    \caption{Performance after 20 epochs of fine-tuning on DeepMIMO as a function of the fine-tuning dataset size, for $\textbf{n}=[1,8,8]$, $-\log\sigma^2=13$, and $m=10^{6}$.}
    \vspace{-15pt}
    \label{fig:NumSample}
\end{figure}
To evaluate the efficacy of our data augmentation technique during fine-tuning, we compared sum-rate performance against the number of samples needed for fine-tuning in the deployment scenario, as illustrated in Figure~\ref{fig:NumSample}. Remarkably, our approach achieves $92\%$ of near-optimal sum-rate with just $20$ samples, outperforming \gls{MAML}, which requires $1000$ samples. Both Deep-All and Rnd-initialized, even with $10^{6}$ samples, fail to reach the performance of SAGE-HB-Net achieved with only $20$ samples. Moreover, we notice that the DNN overfits the training data if we do not use data augmentation for $|\mathcal{D}|$ is less than $10^6$. This shows the importance of data augmentation for adapting to unknown environments. In the cases when $|\mathcal{D}|=10$ or $15$, as the maximum value of ($m$) is less than $10^{6}$, then we sample the user’s CSI with replacement, which acts as a regularizer and prevents overfitting.
It is crucial to note that, despite leveraging unsupervised learning that solely relies on CSI for DNN training, acquiring each sample in the deployment scenario demands time, resources, and energy for channel estimation, highlighting not only the computational efficiency of our method but also its practical advantage in reducing the time required for data collection and model fine-tuning in real-world mMIMO environments.

\vspace{-5pt}
\section{Conclusion} \label{Sec:conclusion}
The promise of DL in mMIMO systems for beamforming faces significant challenges in adapting to real-world scenarios. These challenges include the need for extensive data for adaptability, generalization, and fine-tuning issues. In this paper, we have shown how to configure MLDG for HBF and enable it to generalize to various antenna configurations, path loss exponent factors, and BS heights. Our proposed approach combines MLDG with innovative data augmentation techniques during the fine-tuning process. Our approach not only accelerates the adaptation of DL-based beamforming solutions to new channel environments but also dramatically reduces the data requirements for fine-tuning. This advancement enhances the practicality and efficiency of conventional DL-based beamforming solutions for mMIMO systems. Our simulation results provide strong evidence of the effectiveness of our approach.

\vspace{-3pt}
 \section*{Acknowledgement}
 This work was supported by Ericsson - Global Artificial Intelligence Accelerator AI-Hub Canada in Montr\'{e}al and jointly funded by NSERC Alliance Grant 566589-21 (Ericsson, ECCC, Innov\'{E}\'{E}).

\vspace{-10pt}
\bibliographystyle{IEEEtran}
\bibliography{ICMLCN_2024.bib}

\end{document}